\documentclass[twocolumn,english,superscriptaddress,noeprint,prl,longbibliography]{revtex4-1}
\usepackage{amsmath}
\usepackage{amssymb} %
\usepackage{braket}

\usepackage{dsfont}
\newcommand{\idop}{\mathds{1}}  

\usepackage{hyperref}
\hypersetup{
	colorlinks=true,
	linkcolor=blue,
	filecolor=blue,
	citecolor = black,      
	urlcolor=cyan,
}

\usepackage{catchfilebetweentags}
\newif\ifsupp
\supptrue

\def\rmi{\mathrm{i} }
\def\nn{\nonumber}
\def\Ezero{\Delta}

\newcommand{\rs}{\rm \scriptscriptstyle}
\usepackage{xspace}
\def\hc{\text{h.c.}\xspace}

\newcommand{\figref}[1]{Fig.~\ref{#1}}

\newcommand{\remove}[1]{[REMOVE:\textcolor{gray}{#1}]}
\renewcommand{\remove}[1]{}

\usepackage{graphicx}
\usepackage{physics}

\def\beq{\begin{equation}}
\def\eeq{\end{equation}}
\def\beqa{\begin{eqnarray}}
\def\eeqa{\end{eqnarray}}
\def\down{\ensuremath{\downarrow}}
\def\up{\ensuremath{\uparrow}}
\renewcommand{\paragraph}[1]{{\it #1.---}}

\begin{document}
\title{
Circuit Quantum Electrodynamics in Hyperbolic Space:\\ From Photon Bound States to Frustrated Spin Models 
}
\date{\today}
\author{Przemyslaw Bienias}
\email{bienias@umd.edu}
\affiliation{Joint Quantum Institute, University of Maryland, College Park, MD 20742, USA}
\affiliation{Joint Center for Quantum Information and Computer Science, NIST/University of Maryland, College Park, Maryland 20742, USA}

\author{Igor Boettcher}
\affiliation{Department of Physics, University of Alberta, Edmonton, Alberta T6G 2E1, Canada}
\affiliation{Theoretical Physics Institute, University of Alberta, Edmonton, Alberta T6G 2E1, Canada}

\author{Ron Belyansky}
\affiliation{Joint Quantum Institute, University of Maryland, College Park, MD 20742, USA}
\affiliation{Joint Center for Quantum Information and Computer Science, NIST/University of Maryland, College Park, Maryland 20742, USA}

\author{Alicia J. Koll\'{a}r}
\affiliation{Joint Quantum Institute, University of Maryland, College Park, MD 20742, USA}

\author{Alexey V. Gorshkov}
\affiliation{Joint Quantum Institute, University of Maryland, College Park, MD 20742, USA}
\affiliation{Joint Center for Quantum Information and Computer Science, NIST/University of Maryland, College Park, Maryland 20742, USA}

\begin{abstract}
Circuit quantum electrodynamics is one of the most promising platforms for efficient quantum simulation and computation. 
In recent groundbreaking experiments, the immense flexibility of superconducting microwave resonators was utilized to realize hyperbolic lattices that emulate quantum physics in negatively curved space.
Here we investigate experimentally feasible settings in which a few superconducting qubits are coupled to a bath of photons evolving on the hyperbolic lattice. 
We compare our numerical results for finite lattices with analytical results for continuous  hyperbolic space on the Poincar\'{e} disk. 
We find %
good agreement between the two descriptions in the long-wavelength regime.
We show that photon-qubit bound states have a curvature-limited size. 
We propose to use a qubit as a local probe of the hyperbolic bath, for example by measuring the relaxation dynamics of the qubit.
We find that, although the boundary effects strongly impact the photonic density of states, the spectral density is well described by the continuum theory.
We show that interactions between qubits are mediated by photons propagating along geodesics. 
We demonstrate that the photonic bath can give rise to geometrically-frustrated hyperbolic quantum spin models with finite-range or exponentially-decaying interaction.
\end{abstract}
\maketitle

One of the greatest challenges of modern physics is to formulate a consistent theory that unifies general relativity and quantum mechanics. A possible way to %
shed light on this problem is to study %
well-controlled table-top quantum simulators that mimic curved geometries~\cite{Leonhardt2000,Genov2009,Smolyaninov2010,Bekenstein2015,Garay2000,Fedichev2003,Chang2007,Sheng2013,Lahav2010,Hu2019a}. %
Lattices of microwave resonators in circuit quantum electrodynamics (QED) emerged recently as a particularly promising platform~\cite{Houck2012,Fitzpatrick2017,Anderson2016}.
The high control and flexibility of this system make it possible to incorporate spatial curvature in different models with  strong {quantum} effects.  
Previous studies focused on understanding the properties of photons living on the hyperbolic lattice~\cite{Kollar2019,PhysRevX.10.011009,Boettcher2020,Yu2020a,brower2019lattice,PhysRevD.102.034511,maciejko2020hyperbolic,zhang2020efimov,PhysRevA.103.033703,zhu2021quantum,boettcher2021crystallography}.
In this Letter, we study the impact of negative curvature on various observables of a hybrid system consisting of qubits and photons on a hyperbolic lattice.

For decades, the spectra of hyperbolic graphs have been studied by mathematicians and computer scientists due to their unusual properties~\cite{Woess1987,Sunada1992,Floyd1987,Bartholdi2002,Strichartz1989,Agmon1986,Krioukov2010,Boguna2010}.
Recently, hyperbolic
graphs have also attracted the attention of the quantum error correction community \cite{pastawski2015holographic,Breuckmann2016,Breuckmann2017,Lavasani2019,jahn2021holographic}.
The hyperbolic spectrum can be probed in experiments through transmission measurements~\cite{Kollar2019}. 
Here, we propose to use qubits to probe the local properties of hyperbolic graphs.

Classical spin models on hyperbolic lattices were studied %
in Refs.~\cite{Rietman1992,Shima2006,Baek2008,Krcmar2008c,Baek2009,Gu2012, Serina2016}.
However, the quantum spin model problem ~\cite{PhysRevLett.111.157201,PhysRevB.87.085107,Daniska2016,Chapman2020} is more challenging due to the  interplay of quantum mechanics with non-commutative geometry and strong geometric frustration. %
In the spirit of quantum simulation, we propose to use a hybrid photon-qubit system to engineer (i) finite-range localized photon-mediated interactions between spins, and (ii) exponentially decaying photon-mediated interactions.
Finite-range spin-spin interactions arise from photon localization within flat-bands~\cite{Kollar2019}. %
Interacting bosons in flat bands were studied in Ref.~\cite{Pudleiner2015}, and BCS theory on a flat band lattice in Ref.~\cite{Miyahara2007}. 
Our platform can be used to explore similar physics in curved space by using spins to induce interactions between photons.
Finally, our work is related to recent studies of  emitters interacting with structured quantum baths,
which feature unconventional non-Markovian effects~\cite{Calajo2016,Shi2016,Shi2018b,Gonzalez-Tudela2017a,Gonzalez-Tudela2017}.
\begin{figure}[t!]
\includegraphics[width=1\columnwidth]{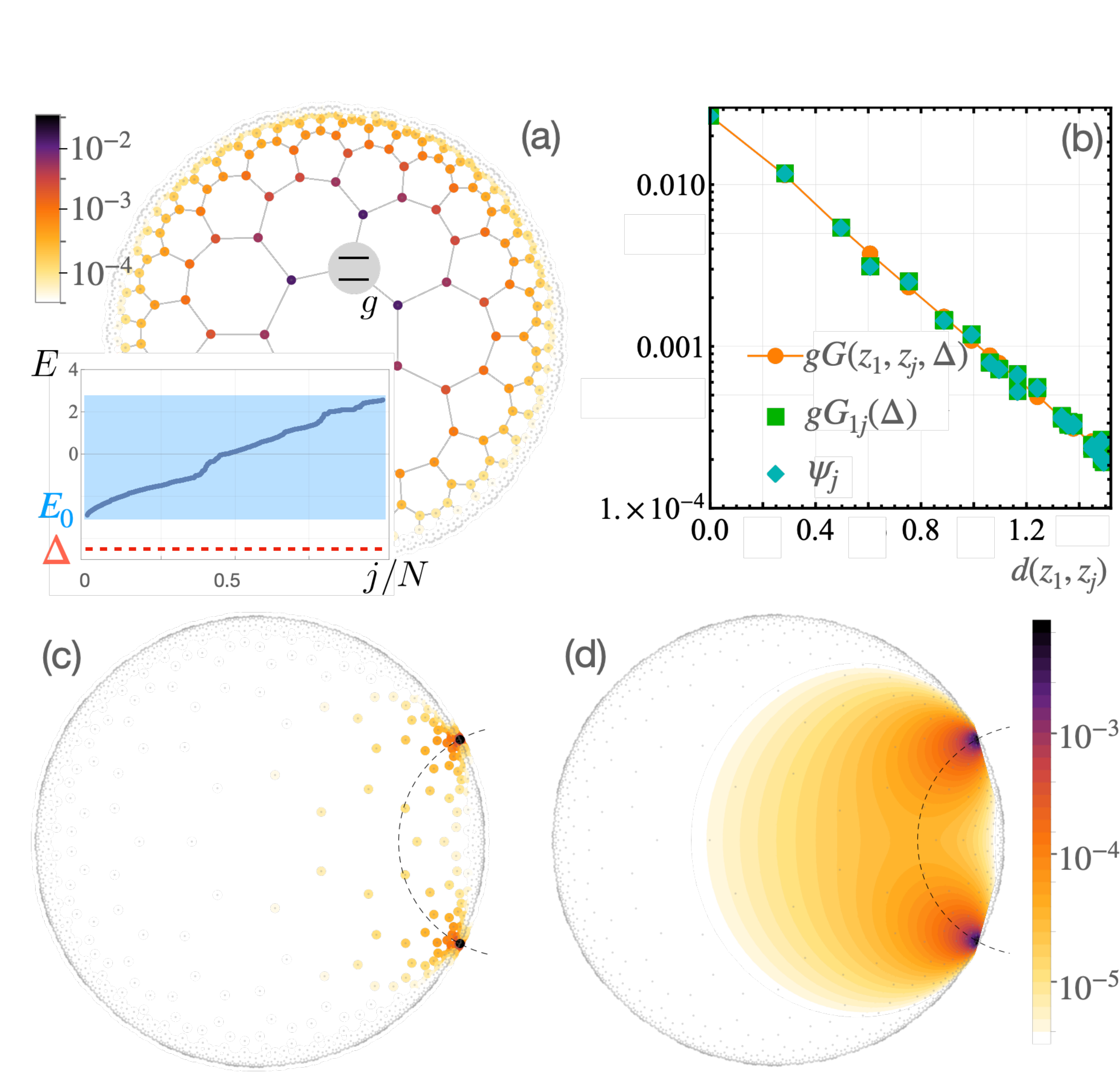}
\caption{
(a) Single-excitation single-qubit bound-state. Color indicates the amplitude of the wavefunction. Gray lines denote the connectivity of photonics sites. One site is coupled via $g$ to a qubit (gray circle). The inset shows the photonic spectrum as a function of the eigenstate  index $j$. 
All results are for  $g=0.05$ and $\Delta=-3.2$ below the band edge at $E_0\approx -2.9$.
(b) Comparison of photonic amplitudes of the lowest eigenstate  as a function of the distance $d(z_1,z_j)$ from the qubit at $z_1$---exact $\psi_j$ (diamonds), perturbative expressions from Eq. (\ref{BS1}) based on discrete (squares) and continuous (dots) Green function.
(c-d) The density of the photonic amplitude of a two-spin single-excitation bound state illustrating the photon-mediated interactions between spins on the graph (c) and in the continuum (d).
We see that photons follow the shortest path between two spins (positioned at two black dots in (c))---the geodesic shown as a (dashed) semi-circle on the Poincar\'{e} disk.
}
\label{fig1}
\end{figure}

\paragraph{System}%
We study photons on a hyperbolic lattice $\mathcal{G}$ coupled to qubits,
where photon dynamics is modeled by a tight-binding Hamiltonian, and qubits at positions $i$ correspond to local spin-$1/2$ operators $\sigma^+_i\sigma^-_i = \ketbra{\up}{\up}_i$. The full Hamiltonian in the rotating-wave approximation and rotating frame %
is given by 
\begin{eqnarray}
\hat{H}&=& \Ezero \sum_{i\in\mathcal{S}} \ketbra{\up}{\up}_i
 +g\sum_{j\in\mathcal{S}}\Bigl(\sigma^+_ja_j+\hc\Bigr)+\hat{H}_{\rs ph},\\
\hat{H}_{\rs ph}&=& - t\sum_{<ij>\in \mathcal{G}}a_i^\dagger a_j, %
\end{eqnarray}
with $a_i^\dagger$ the photon creation operator, $g$ the coupling between photons and qubits, and $\Delta$ the difference between the qubit frequency and the frequency of a photon on a single site. The set $\mathcal{S}$ comprises the qubit sites and $\mathcal{G}$ the hyperbolic lattice. For concreteness, in the following we set the hopping $t=1$.

The spectrum of $\hat{H}_{\rs ph}$ is bounded from below and above \cite{kollar2019line,kollar2021gap} [see Fig.~\ref{fig1}(a)], and hence the spectrum of $\hat{H}$ consists of scattering eigenstates together with localized photon-qubit bound states~\cite{Sundaresan2019,Calajo2016,Munro2017,Liu2017a,Munro2017,Hung2016,Douglas2016}.
For a single qubit at position $i$, the single-excitation bound-state energy $E_{\rm B}$ is given by the solution of
\begin{equation}
E_{\rm B}=\Ezero + g^2G_{ii}(E_{\rm B}),
\label{eq:EBS}
\end{equation}
where $G_{ij}(\omega)=(\omega-\hat{H}_{\rs ph})^{-1}_{ij}$ is the photonic Green function. %
We denote the lowest eigenvalue of $\hat{H}_{\rm ph}$ or lower band edge (LEBE) by $E_0<0$.
Equation (\ref{eq:EBS}) always permits two solutions outside the photonic band and we focus in the following on the lower bound state with $E_{\rm B}<\Ezero$. For weak coupling ($g\ll |E_0-\Delta|$), we have
$E_{\rm B}\approx\Ezero+g^2 G_{ii}(\Ezero)$, and the bound-state wavefunction consists mostly of the spin component such that
\beqa
\ket{\psi_{\rm B}}\approx \ket{\up,0} + g\sum_{j\in \mathcal{G}} G_{ij}(\Ezero)a_j^\dagger\ket{\down,0}.
\label{BS1}
\eeqa
When the experimentally relevant energies are close to the LEBE, $\omega-E_0\lesssim t$,
we can capture the photonic part by a continuum model \cite{Boettcher2020} on the Poincar\'{e} disk with metric $\mbox{d}s^2=(2\kappa)^2(\mbox{d}x^2+\mbox{d}y^2)/(1-r^2)^2$ and curvature radius $\kappa=1/2$ \cite{Cannon}.
The finite hyperbolic lattice is mapped to a hyperbolic disk of radius $L<1$, where $L=\sqrt{N/(N+N_0)}$  for $N$ sites with $N_0$ a constant~\cite{Boettcher2020}, so that  
$L\to 1$ for large lattices.  For concreteness, we consider the $\{7,3\}$  lattice based on regular heptagons with coordination number 3, with $N_0=28$ and lattice constant $h=0.276$. However, our results apply \emph{(i)} to other hyperbolic $\{p,q\}$ lattices by substituting the corresponding value of $h$, and \emph{(ii)} to line-graphs of $\{p,q\}$ lattices \cite{kollar2019line} in the long-wavelength regime.  
 We denote the hyperbolic distance by $d(z,z')=\kappa\ \text{arcosh}( 1+\frac{2|z-z'|^2}{(1-|z|^2)(1-|z'|^2)})$. The number $d(z_i,z_j)$, intuitively, quantifies the number of hops needed to get from $z_i$ to $z_j$ on the graph.

The photon spectrum is continuous on the Poincare\'{e} disk and given by $E_{\textbf{k}}=E_0+\frac{1}{M}|\textbf{k}|^2$, with momentum $\textbf{k}$, effective photon
mass $M=\frac{4}{3h^2}$~\cite{Boettcher2020}. 
The bound-state condition~\eqref{eq:EBS}
becomes %
$E_{\rm B}=\Ezero + g^2G_\Lambda(z_i,z_i,E_{\rm B})$, with $G_\Lambda(z,z',\omega)$ the continuum approximation of the photon Green function. 
The subscript $\Lambda$ indicates the need to introduce a large-momentum
cutoff $\Lambda\propto h^{-1}$, because the continuum Green function is not well-defined for $z=z'$~\cite{Boettcher2020}. This is analogous to the well-known regularization of bound states for parabolic bands in two Euclidean dimensions \cite{Randeria1989,SchmittRink}. 
The value of $\Lambda$ can be fixed through a renormalization condition, %
yielding $\Lambda \simeq 3\,h^{-1}$~\cite{supplementHyperFew}.
The bound-state wavefunction for a qubit at $z_i$, for arbitrary $L<1$ and energies close to the LEBE, is 
\begin{equation}
	\ket{\psi_B} \propto \Bigl(\sigma^+_i-\int \frac{\mbox{d}^2z}{(1-|z|^2)^2} u(z)a^\dagger(z)\Bigr) \ket{\downarrow,0},
\end{equation}
with $u(z) = M \tilde{g}G_\Lambda(z_i,z,E_{\rm B})$ and $\tilde{g}=\sqrt{\pi/28}g$. In Fig.~\ref{fig1}(b), we show the photonic amplitude $|u(z)|$ of the bound-state wavefunction %
using both the continuum expression for $G_\Lambda(z_i,z_j,E_{\rm B})$ from Ref.~\cite{Boettcher2020} and lattice Green function $G_{ij}(E_{\rm B})$, which both agree well with the exact result.

Setting $L=1$ in the continuum model, i.e. considering an infinite system, often leads to simple analytical formulas, but assumes the absence of a system boundary. Since boundary effects are not subleading in hyperbolic space, results obtained for $L=1$ can differ \emph{qualitatively} from those for any $L<1$. In this work, we
are primarily interested in contributions of weakly coupled spins located in the bulk, sufficiently far away from the boundary. We find that some observables are well-captured by the $L=1$ limit, whereas others need to be computed for $L<1$. The continuum Green function for $L=1$ reads 
\begin{align}
G_{\Lambda}(z_i,z_i,E_{\rm B})=\frac{\pi}{28} \int_{k\leq \Lambda} \frac{\mbox{d}^2k}{(2\pi)^2} \frac{\tanh(\pi k/2)}{E_{\rm B}-E_0-\frac{1}{M} k},
\label{BS2}
\end{align}
where $k=|{\bf k}|$.
The tanh-term in the numerator \cite{helgason2006non} is due to the negative curvature of space. 
 In practice, $|E_{\rm B}-E_0|\gg M^{-1}$ because of the large value of the mass, and we can neglect the tanh-factor to approximate $G_\Lambda(z_i,z_i,E_{\rm B})\approx \frac{M}{112}\ln(|E_B-E_0|M/\Lambda^2)$.

\paragraph{Curvature-limited correlations}%
The continuum approximation enables us to analytically quantify 
the size of the single-particle bound state. For this purpose, we expand~\cite{supplementHyperFew} the Green function %
for large hyperbolic distance
$d(z,z')\gg h$, leading to %
\beq 
G_\Lambda(z,z',{\omega})\approx G_\Lambda^{(0)}(\omega) %
\exp\Bigl(-\frac{d(z,z')}{2\xi({\omega})}\Bigr),
\label{EqDecay}
\eeq
which confirms the exponential decay shown in Fig.~\ref{fig1}(b).
 The correlation length $\xi$ depends on the system parameters through the frequency ${\omega}$, and in the following we neglect  %
a weak residual dependence on $z$ and $z'$ from boundary effects. %
In the continuum approximation and in the limit $L\to 1$, the lower edge of the photonic band is located at $E_0 = -3+ \frac{1}{M}$~\cite{Boettcher2020}.  %
 As $\omega$ approaches $E_0$ from below, for qubits coupled to a Euclidean lattice, the correlation length diverges as $\xi \propto (E_0-\omega)^{-1/2}$ ~\cite{Sundaresan2019,Shi2016}. In stark contrast, on the hyperbolic lattice, correlations are cut off by the curvature radius and $\xi\leq \kappa$ remains finite. In particular, for $L=1$ and %
 $\omega< E_0$, we  find~\cite{supplementHyperFew} that
\beq
\xi \approx %
\frac{\kappa}{1 + \sqrt{M(E_0-\omega)}}
\label{xiLabel}.
\eeq
\paragraph{Spin relaxation and photonic density of states (DOS)}%
We propose a local probe---an excited qubit with frequency within the band---to measure properties of hyperbolic graphs.
For very weak coupling $g$, one can couple to only a few eigenstates and extract the spectral properties from the time dependence of the excited state population.
On the other hand, by using larger $g$, such that a qubit couples to many states, the dynamics of the initially excited spin corresponds to the exponential decay %
governed by the graph spectrum. We concentrate on the latter in the following. 

The spontaneous emission from the qubit
can be described by a Markovian Lindblad master equation. The decay rate is given by $\Gamma = j(\Delta)$,
where $j(\omega)=2\pi\sum_{\textbf{k}}\abs{g_{\textbf{k}}}^2\delta(\omega-\omega_{\textbf{k}})$ is the spectral function. In the low-energy continuum approximation we have $g_{\textbf{k}}=g\sqrt{\frac{\pi}{28}}|\psi_{\textbf{k}}(z_i)|$ with $\psi_{\textbf{k}}(z)$ the eigenfunctions of the hyperbolic Laplacian%
~\cite{supplementHyperFew} and $\omega_{\textbf{k}}=E_0+\frac{1}{M}|\bf k|^2$. Furthermore, for $L=1$, we have
\begin{equation}
j_{L=1}(\omega)=\frac{\pi M}{56}g^2\tanh\left(\frac{\pi\sqrt{(\omega-E_0)M}}{2}\right).
\label{eq:jL}
\end{equation}
We see that, due to the tanh factor, $j(\omega)$ is qualitatively different in curved space than in
2D Euclidean space where, for quadratic dispersion, $j(\omega)$ lacks this factor and is thus constant. 
However, the range of energies for which curved and flat space differ is restricted to a narrow energy range $\lesssim\delta E_{L=1} = 1/M\ll 1$ close to the LEBE.
Note that, for $L=1$ and within the continuum approximation, $j(\omega)$ is directly proportional to the DOS $\rho$, $\rho_{L=1}=N\eta \tanh(\frac{\pi}{2}\sqrt{(\omega-E_0)M})$ with $\eta=\frac{M}{112}$, as the energy dependence from $g_{\textbf{k}}$ drops out in the angular average~\cite{supplementHyperFew}.

Close to the LEBE, the photonic spectrum can be computed from the hyperbolic Laplacian $\Delta_g$ for any $L<1$ using Dirichlet boundary conditions. The DOS of eigenvalues $\varepsilon$ of the Laplacian follows Weyl's law $\rho_W(\varepsilon) \sim \frac{\text{area}}{4\pi}-\frac{\text{circ}}{8\pi \sqrt{\varepsilon}}+\mathcal{O}(\varepsilon^{-1})$ ~\cite{pathria,Ivrii1980,Osgood1988,Aydin2018}, %
with $\text{area}=\pi L^2/(1-L^2)$ and $\text{circ}= 2\pi L/(1-L^2)$ the area and circumference of the finite hyberbolic disk, respectively. Using $\varepsilon=M(\omega-E_0)$, we arrive at %
\begin{align}
\rho_{\rm W}(\omega) \approx \frac{L^2/h^2}{3(1-L^2)}-\frac{L/h}{2\sqrt{3}(1-L^2)}\frac{1}{\sqrt{\omega-E_0}}.
\label{EQweyl}
\end{align}
Note that the leading, constant term is characteristic for parabolic bands in two dimensions, but the subleading correction is important even in the large-system limit---which is in dramatic contrast to flat space. Importantly, setting $L=1$ in $\rho_{\rm W}$ does not reproduce $\rho_{L=1}$.

\begin{figure}[t!]
\includegraphics[width=1\columnwidth]{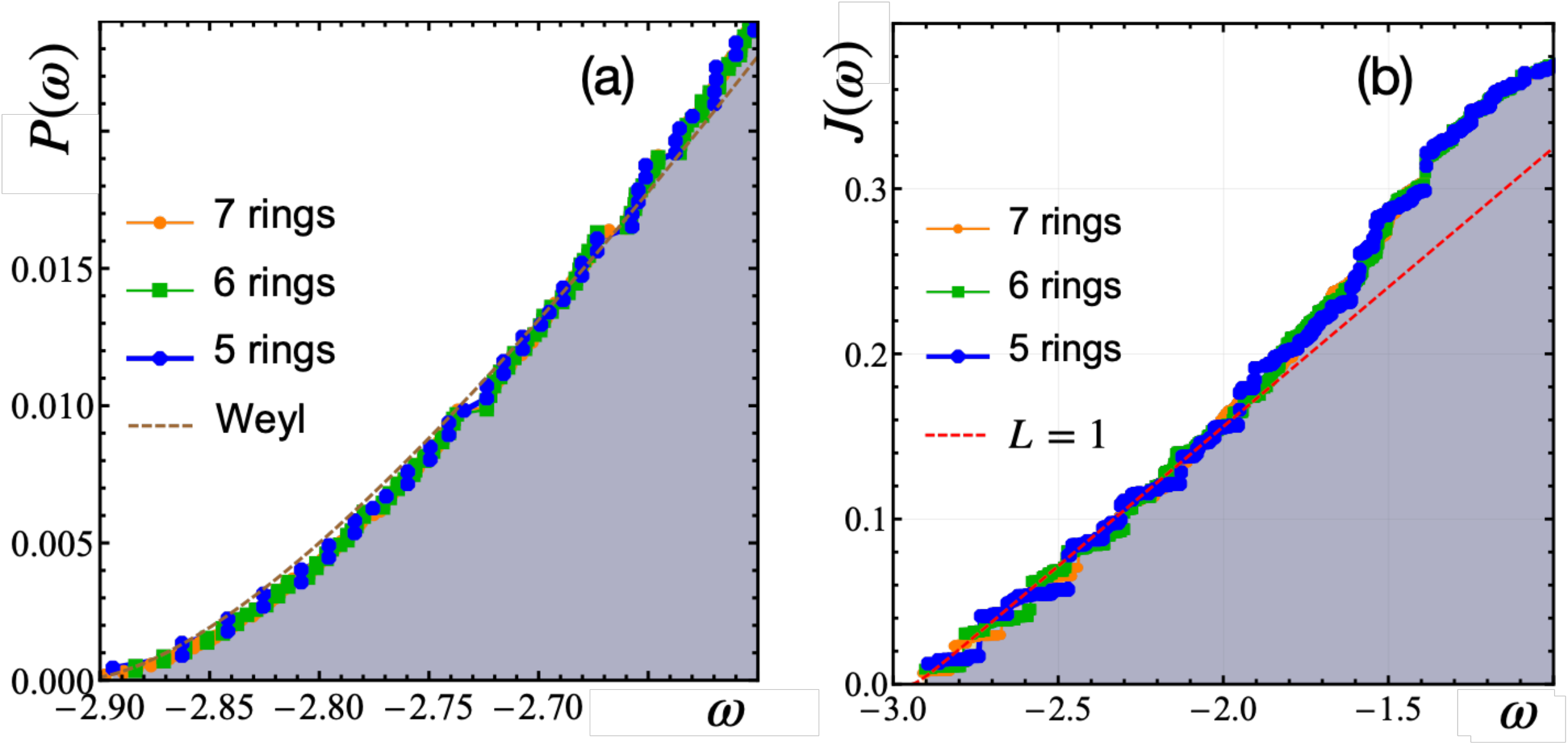}
\caption{
Comparison of spectral properties from analytic limits to numerics of different-sized graphs in the long-wavelength limit near the LEBE. Graphs are characterized by the number $\ell$ of concentric rings of heptagons, where $\ell=1$ corresponds to a single heptagon.
(a) The cumulative DOS $P(\omega)$ has da\-tapoints nearly overlapping for different $\ell$.
(b) The cumulative $J(\omega)$ is less smooth, but asymptotically ($\ell\rightarrow \infty$) approaches %
$J_{L=1}(\omega)$: The root-mean-square error with respect to $J_{L=1}$ for $\omega<-2$ is 0.0066, 0.0052, and 0.0044 for $\ell=5,6$, and $7$, respectively.
}\label{fig:rho_j}
\end{figure}

In \figref{fig:rho_j}(a), we plot the exact normalized cumulative DOS, $P(\omega)=\frac{1}{N}\int^\omega_{E_0} \text{d}\nu\, \rho(\nu)$,
from which we see that $P$ shows Weyl scaling and does not change with system size. We confirm that the 
second term in
Weyl's law (\ref{EQweyl}) is non-negligible in a wide range of energies, $\delta E_{\rm W}$, which 
is much greater than $\delta E_{L=1}$. The neglect of boundary effects in $\rho_{L=1}$ does not reproduce the lattice DOS.
Note that the leading term in $\rho_W$ is the same as the  %
large-$\omega$ value of $\rho_{L=1}(\omega)$, which is equal to $N\eta$. %

Figure \ref{fig:rho_j}(b) shows the cumulative spectral function  $J(\omega)=\int^\omega_{E_0} \text{d}\nu\, j(\nu)/g^2$. Close to the band edge, we observe that $J(\omega)$ with increasing $\ell$ is well-approximated by a constant $j(\omega)$. 
Hence, $J(\omega)$ qualitatively agrees with %
$j_{L=1}(\omega)$, because both lead to a nearly constant value of $j(\omega)$. (Except 
in a frequency window of size
$\delta E_{L=1}\ll 1$ near the band edge, which we cannot resolve numerically with finite lattices, and thus cannot resolve the tanh-factor in Eq. (\ref{eq:jL}).) We conclude that, since $j(\omega)$ is a \emph{local} quantity, it is only weakly influenced by the boundary physics and can be described by the continuum $L=1$ theory, which is an important and useful result.

\begin{figure}[t!]
\includegraphics[width=1\columnwidth]{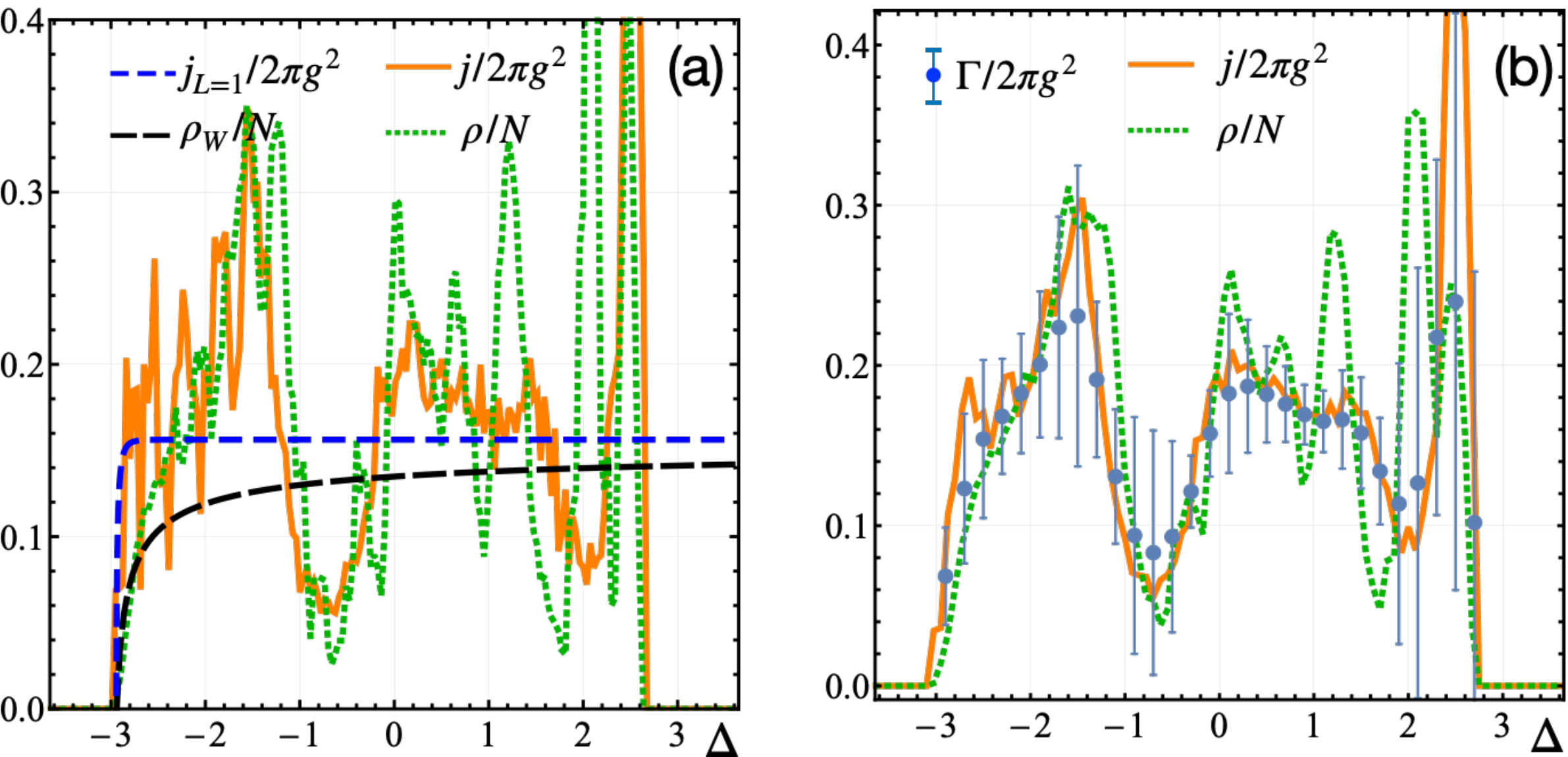}
\caption{
 (a) Comparison of analytical expressions and numerical results (with the bin size $\Delta \omega=0.15$) for $j$ and $\rho$.
 (b)
 Comparison of the numerical (with $\Delta \omega=0.3$) $j$ and $\rho$  with the fitted $\Gamma$ within time $\in [0,15]$ (error bars correspond to the standard deviation) for  $g=0.3$, $\ell=7$.
}
\label{fig:decay}
\end{figure}

The presence of the lattice leads to a significant difference between $\rho$ and $j$ for large $\Delta$ away from the LEBE [see \figref{fig:decay}(a)], whereas they are directly proportional in the low-energy continuum approximation for $L=1$.
We numerically extract the decay rate $\Gamma$ on the lattice from the dynamics of an initially excited spin with no photons.
We find that
$\Gamma \propto j(\omega)$ rather than $\propto\rho$ [see \figref{fig:decay}(b)], however, with significant error bars for parameters accessible to us.
Due to the limited lattice sizes accessible in numerics,
we analyzed $g=0.3$ to ensure that the qubit is coupled to many photonic modes. %
The deviation between $\Gamma$ and $j$ is caused by (i) the finite number of states we couple to, (ii) edge effects such as reflection from the boundary, and (iii) effects beyond Fermi's golden rule due to $g$ being comparable to the hopping strength $t=1$.

\begin{figure}[t!]   
\includegraphics[width=1\columnwidth]{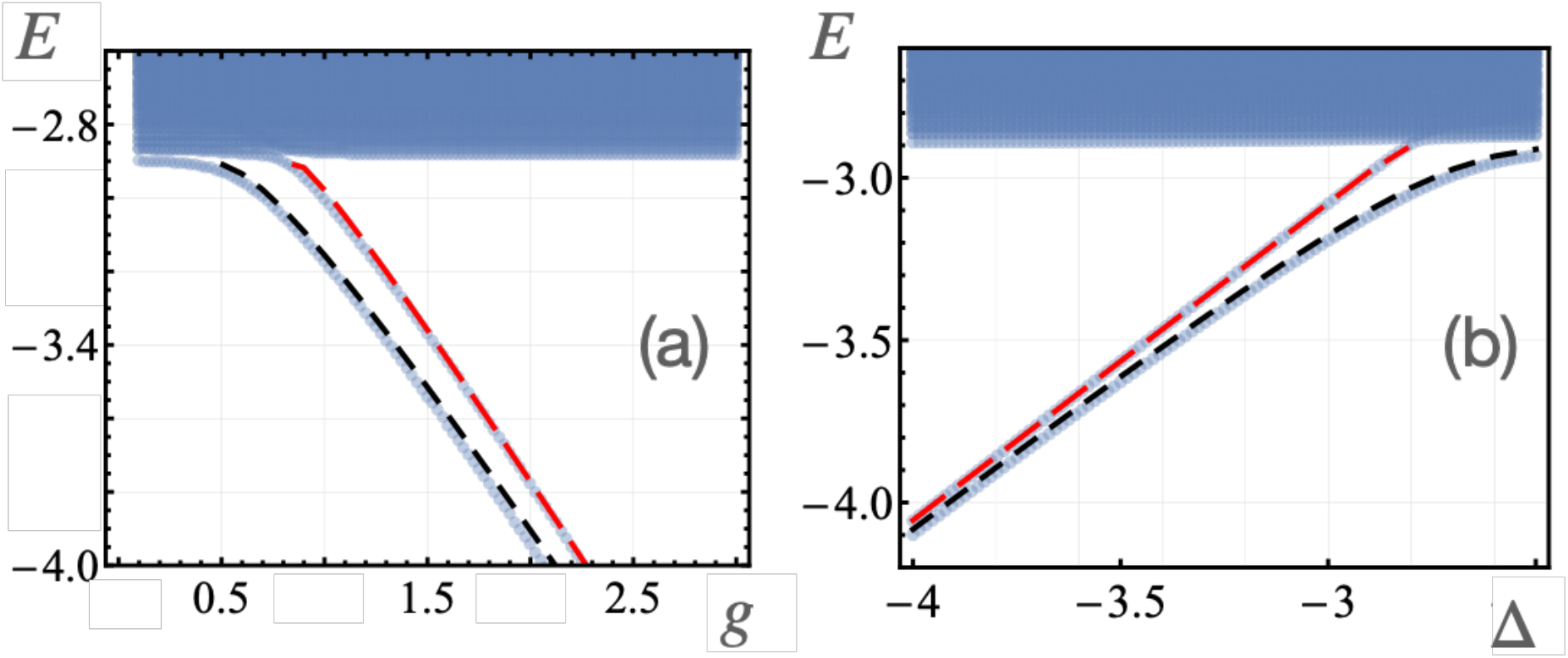}
\caption{
Energy spectrum (a) as a function of $g$ for $\Delta=-2.5$ and next-nearest neighbors separated by $z=0.5$, and (b) as a function of $\Delta$ for $g=0.5$ and nearest neighbors separated by $z=h=0.276$.
All lattice results %
are for $\ell=6$, which we compare with symmetric (lower black dashed curve) and anti-symmetric (upper red dashed) two-qubit bound states.
}
\label{figBS_2qubits}
\end{figure}

\paragraph{Single-excitation bound state for two qubits}%
We now consider two qubits located at positions $z_1$ and $z_j$, each with frequency $\Ezero$. 
The energies of two single-excitation bound states ($E_{\rm B}^\pm$) 
are given by the solutions of 
\begin{align}
\label{eq:2spin-boundstates}
E_{\rm B}^\pm-\Ezero-\Sigma_{11}(E_{\rm B}^\pm)\mp \Sigma_{1j}(E_{\rm B}^\pm)=0,
\end{align}
with $\Sigma_{ij}(\omega) =  g^2G_{ij}(\omega)$.
We find good agreement between the solution using the continuum Green function $G_\Lambda(z_i,z_j,\omega)$ and the results using the lattice Green function $G_{ij}(\omega)$ (see \figref{figBS_2qubits}). 
The bound-state wavefunctions in the continuum are 
\beqa
	\ket{\psi_{\rm B}^\pm} &&= 
	c_{\pm}\Big{[} (\sigma^+_1\pm\sigma^+_j)
	\\
	-&&\frac{4\tilde{g}}{3th^2}\int_z \left[G_\Lambda(z,z_1;E_{\rm B}^\pm)\pm G_\Lambda(z,z_j;E_{\rm B}^\pm)\right]a^\dagger(z)\Big{]}\ket{\downarrow\downarrow0}.\nn
	\label{PsiBpm}
\eeqa
In Fig.~\ref{fig1}(c-d), we plot
the photonic density for the lowest (i.e.~symmetric) bound state $\ket{\psi^+_{\rm B}}$ as a function of position $z$. On a lattice, $z$ takes only discrete values $z_i$ and  $n_{\rm ph}(z_i)$ is given by $g^2 \sum_{w,w'\in\{1,j\}}G_{wi}(E_{\rm B})G_{w'i}(E_{\rm B})$.  %
We see that the photons mediating the interactions follow a geodesic---the shortest path between the two spins.

\paragraph{Effective spin models}%
The coupling of spins to the hyperbolic photonic bath leads to an effective spin-spin interaction that changes as the spin frequency is tuned. %
For spin frequencies satisfying  $E_0-\Ezero\gg g$, we can integrate out the hyperbolic photons and arrive at the effective spin-spin Hamiltonian
\beqa
\hat{V} ={g^2}\sum_{ij}G_{ij}(\Ezero) \sigma_i^-\sigma_j^+ %
\eeqa
describing flip-flop interactions. %
Here terms with $i=j$ introduce an on-site energy shift for the qubits. For $i\neq j$, we can use the continuum approximation for the Green function $G_{ij}(\omega)$. Equation (\ref{EqDecay}) reveals that the interaction decays exponentially with a correlation length $\xi\leq \kappa=1/2$ that depends on $\Ezero$.

\begin{figure}[t!]   
\includegraphics[width=.82\columnwidth]{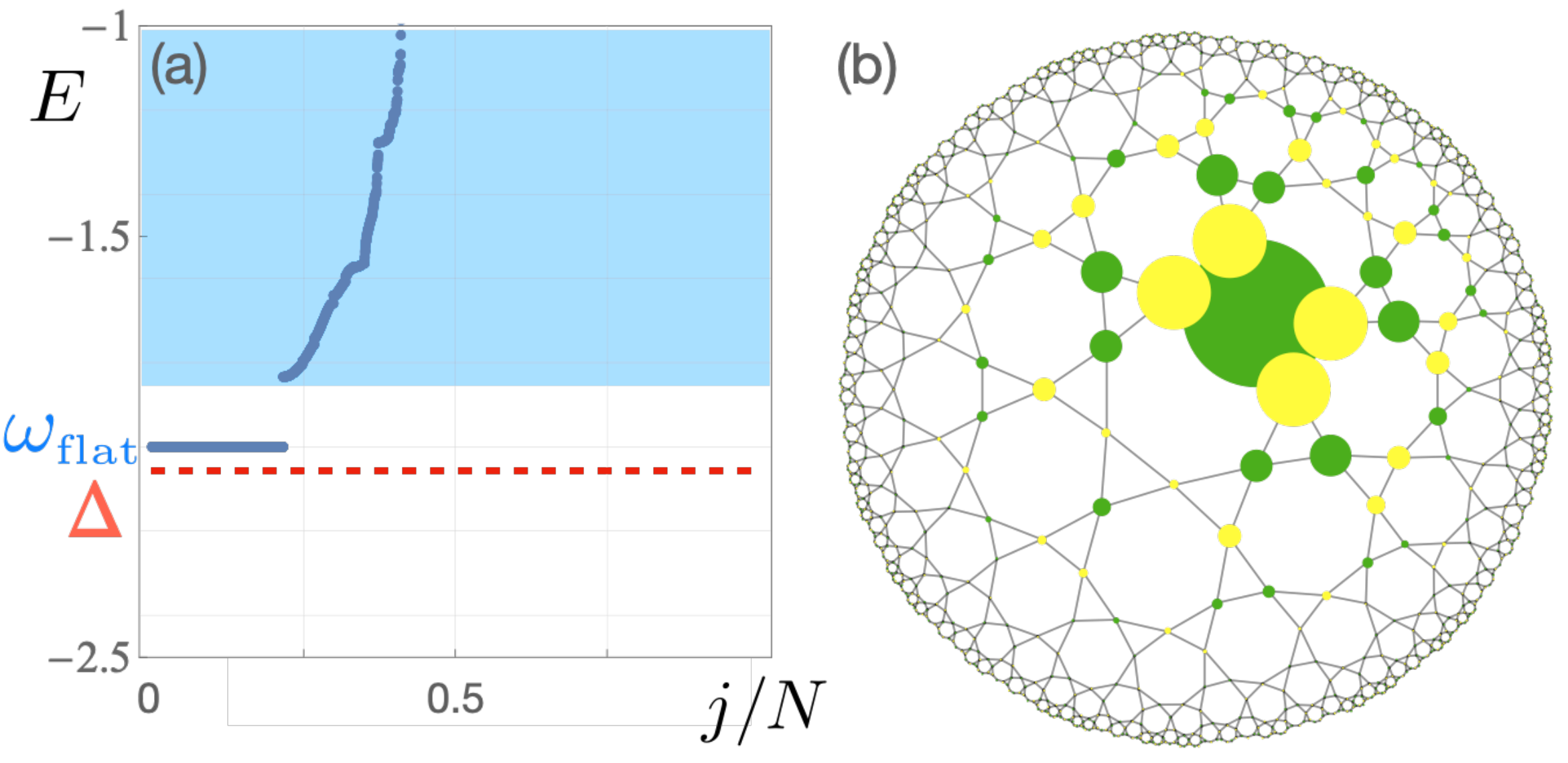}%
\caption{Spin model in the vicinity of the flatband for the heptagonal kagome-like line graph with $t=-1$~\cite{Kollar2019}. (a) The photonic spectrum as a function of the eigenstate index $j$. (b) We plot the interaction strength as a function of qubit position, assuming the position of the second qubit to be where the largest dot is. The dot color denotes the sign of the interactions (green/yellow is $+/-$), whereas its diameter is directly proportional to the interaction strength. 
Note that the spin model is highly frustrated. 
}
\label{fig:flat_interaction}
\end{figure}

For simplicity of presentation, up to now, we focused on hyperbolic $\{p,q\}$ graphs. %
Conveniently, their
long-wavelength physics is the same as that of their line graphs~\footnote{For line graphs, for $t<0$, the long-wavelength part of the spectrum corresponds to high energies.},  which naturally appear in cQED experiments~\cite{Kollar2019}.
Line graphs, additionally, feature a flatband of localized states at energy $\omega_{\rm flat}$, which is near the LEBE for $t<0$~\cite{Kollar2019}. 
In particular, for  lattices based on polygons with an odd number of vertices, these flatbands are gapped [see Fig.~\ref{fig:flat_interaction}(a)].
Due to the gap, we can choose the qubits to be coupled effectively only to the flatband. In that case, because flaband eigenstates have support only on two neighboring polygons~\cite{Kollar2019}, the resulting   photon-mediated interactions between the spins are strictly finite-range.
The spin Hamiltonian, in terms of the wavefunctions $\phi_k$ describing the localized states in the flatband, is given by
\beqa
\hat{V} =\frac{g^2}{\Ezero-\omega_{\rm flat}} \sum_{k,ij} \phi_k^*(z_j)\phi_k(z_i) \sigma_i^-\sigma_j^+. %
\eeqa
Figure \ref{fig:flat_interaction} illustrates that the interactions are of finite range, with the sign of the interactions oscillating rapidly with distance. 
Together with possible geometric frustration, this leads to a highly frustrated spin model, which may lead to exotic spin-liquid phases in cQED experiments.

\paragraph{%
Outlook}%
Using our formalism, it is exciting to study the impact of spatial curvature on fractional quantum Hall phases~\cite{Yu2020a} and on the creation and detection of entanglement.
By coupling a single qubit to multiple resonators, one can realize so-called giant atoms~\cite{Guo2020j}, which might exhibit nonlocal curvature effects in hyperbolic lattices.
Another promising direction is to study the interplay of curved-space effects and frustration in %
(i) spin models based on the interactions we have derived and (ii) bosonic models with spin-induced interactions between photons. Naturally appearing disorder in the experiments could lead to novel physics~\cite{Liu2020h} in spin models and bosonic models.  %
Finally, using the flexibility of cQED, we envision the possibility to engineer a quantum simulator of the AdS-CFT correspondence~\cite{Maldacena1999,Witten1998} to shed light on strongly-correlated condensed matter and weakly-interacting gravity problems.
\acknowledgments{
We thank I. Carusotto, A. Deshpande, A. Ehrenberg, A. Guo, A. Houck, R. Lundgren, and M.\ Tran for discussions. 
P.B., I.B, R.B., and A.V.G.\ acknowledge support by AFOSR, AFOSR MURI, NSF PFC at JQI, ARO MURI, DoE ASCR Quantum Testbed Pathfinder program (award No.\  DE-SC0019040), U.S.\  Department of Energy Award No.\ DE-SC0019449, DoE ASCR Accelerated Research in Quantum Computing program (award No.\ DE-SC0020312), and NSF PFCQC program.
I.B.\ acknowledges support from the University of Alberta startup fund UOFAB Startup Boettcher.
R.B.\ also acknowledges support of NSERC and FRQNT of Canada.
}

\bibliography{main.bbl}
\clearpage
\begin{widetext}
\begin{center}
{\Large \centering Supplemental material}
\end{center}
\setcounter{figure}{0}

\makeatletter
\renewcommand{\bibnumfmt}[1]{[S#1]}
\renewcommand{\citenumfont}[1]{S#1}

\setcounter{equation}{0}
\setcounter{figure}{0}
\setcounter{table}{0}

\renewcommand{\thefigure}{S\@arabic\c@figure}
\renewcommand{\thesection}{S.\Roman{section}}
\renewcommand \theequation{S\@arabic\c@equation}
\renewcommand \thetable{S\@arabic\c@table}

In this supplement, we present technical details omitted from the main text. %
In section \ref{sec:Hcont}, we derive the qubit-photon Hamiltonian within the low-energy continuum approximation.
In section \ref{sec:cutoff}, we discuss the ultraviolet cutoff.
In section~\ref{sec:limited}, we derive Eq.~(\ref{xiLabel}) for the curvature-limited correlations in the limit $L=1$.
In section~\ref{sec:bs}, we derive the bound-state equations for a single qubit and for two qubits.
In section~\ref{sec:emission}, we derive the spontaneous emission rate of an excited qubit within the continuum approximation---Eq.~\eqref{eq:jL} from the main text.

\section{Hamiltonian within the continuum approximation \label{sec:Hcont}}
In this section, we derive the qubit-photon Hamiltonian within the low-energy continuum approximation developed in Ref.~\cite{Boettcher2020}.

We embed the hyperbolic lattice into the Poincare disk $\mathbb{D}=\{z\in \mathbb{C},\abs{z}<1\}$. Sums over lattice sites can be approximated by integrals in the continuum using the following relation
\begin{equation}
\sum_i \rightarrow \frac{28}{\pi}\int \frac{d^2z}{(1-\abs{z}^2)^2}.
\end{equation}

The discrete bosonic operator $a_i$ (satisfying $\comm{a}{a^\dagger}=1$) can then be approximated by a continuum field operator
\begin{align}
\label{eq:a-cont-def-comm}
a(z_i) =& \sqrt{\frac{28}{\pi}}a_{i},
\end{align}
which satisfies the commutation relation 
\begin{equation}
\comm{a(z)}{a^\dagger(z')}=(1-\abs{z}^2)^2\delta^2(z-z').
\end{equation}

As shown in Ref.~\cite{Boettcher2020}, the tight-binding Hamiltonian from Eq.~(1) of the main text is well-approximated by
\begin{eqnarray}
\hat{H}_{\rs ph}&=& - t\sum_{<ij>\in \mathcal{G}}a_i^\dagger a_j \rightarrow \int \frac{d^2z}{(1-\abs{z}^2)^2}\ a^\dagger(z)\qty(-3t-t\frac{3}{4}h^2\Delta_g)a(z),
\end{eqnarray}
where $\Delta_g=(1-\abs{z}^2)^2(\partial_x^2+\partial_y^2)$ is the hyperbolic Laplacian.

The full Hamiltonian from Eq.~(1) is then given by 
\begin{eqnarray}
\hat{H}&=& \Ezero \sum_{i\in\mathcal{S}} \ketbra{\up}{\up}_i
 +\tilde{g}\sum_{j\in\mathcal{S}}\Bigl(\sigma^+_ja(z_j)+\hc\Bigr)+\hat{H}_{\rs ph},\\
\hat{H}_{\rs ph}&=& \int \frac{d^2z}{(1-\abs{z}^2)^2}\ a^\dagger(z)\qty(-3t-t\frac{3}{4}h^2\Delta_g)a(z), %
\end{eqnarray}
where $\tilde{g}=g\sqrt{\frac{\pi}{28}}$.

We will also need the Hamiltonian in Fourier space, valid for the infinite system ($L=1$).
In that case, we have the following Fourier transform relations \cite{helgason2006non}:
\begin{align}
\label{eq:Fourier-rels}
a(z) =& \int \frac{d^2k}{(2\pi)^2}\tanh(\frac{\pi k}{2})\psi_K(z)a(K),\\
a(K) =& \int d^2z\frac{1}{(1-\abs{z}^2)^2}\psi_K^*(z)a(z).
\end{align}
Here, $K=k e^{\rmi \beta}$ or $\textbf{k}=k(\cos\beta,\sin\beta)^T$, and
\begin{align}
    \psi_K(z) = \psi_{\textbf{k}}(z)=\Bigl(\frac{1-|z|^2}{|1-z e^{-\rmi \beta}|^2}\Bigr)^{\frac{1}{2}(1+\rmi k)}
\end{align}
is an eigenfunction of the hyperbolic Laplacian with eigenvalue $-(1+k^2)$.
The Fourier transformed operators satisfy 
\begin{equation}
\comm{a(K)}{a^\dagger(K')} = \frac{(2\pi)^2}{\tanh(\frac{\pi k}{2})}\delta^2(K-K').
\end{equation}

The Hamiltonian then becomes 
\begin{eqnarray}
\label{eq:H-cont-K-space-final}
\hat{H}&=& \Ezero \sum_{i\in\mathcal{S}} \ketbra{\up}{\up}_i
 +\tilde{g}\sum_{j\in\mathcal{S}}\int \frac{d^2k}{(2\pi)^2}\tanh(\frac{\pi k}{2})\Bigl(\psi_K(z_j)a(z_j)\sigma^+_j+\hc\Bigr)+\hat{H}_{\rs ph},\\
\hat{H}_{\rs ph}&=& \int \frac{d^2k}{(2\pi)^2}\tanh(\frac{\pi k}{2}) \omega(k)a^\dagger(K)a(K),
\end{eqnarray}
where
\begin{align}
\label{eq:wk-def}
\omega(k) =& -3t+t\frac{3}{4}h^2(k^2+1).
\end{align}

\section{Ultraviolet cutoff \label{sec:cutoff}} 
In this section, we discuss the ultraviolet (UV) cutoff.

Let us first fix the UV cutoff $\Lambda$ in
\begin{align}
G(z,z',\lambda) = - \int_{k\leq \Lambda} \frac{\mbox{d}^2k}{(2\pi)^2} \tanh\Bigl(\frac{\pi k}{2}\Bigr) \frac{\psi_K(z)\psi_K^*(z')}{\lambda-(k^2+1)}.
\end{align}
We insert $z=z'=z_1$ and find
\begin{align}
 G(z_1,z_1,\lambda) = - \int_{k\leq \Lambda} \frac{\mbox{d}^2k}{(2\pi)^2} \frac{\tanh(\frac{\pi k}{2}) }{\lambda-(k^2+1)}.
\end{align} 
Compare this to the continuum approximation result \cite{Boettcher2020}
\begin{align}
 G_{ij}(\omega) =  M \frac{\pi}{28} G\Bigl(z_i,z_j,\lambda=M(\omega+3)\Bigr),
\end{align}
where $M=\frac{4}{3h^2}=17.529 $ for the $\{7,3\}$ lattice. We conclude that the lattice Green function for $L\to 1$ can be approximated by
\begin{align}
 G_{11}(\omega) &= -M \frac{\pi}{28} \int_{k\leq \Lambda} \frac{\mbox{d}^2k}{(2\pi)^2} \frac{\tanh(\frac{\pi k}{2}) }{M(\omega+3)-(k^2+1)}.
\end{align}
To fix $\Lambda$, we insert $\omega=-3$, which is below the band edge and so leads to a finite integral. We define
\begin{align}
 C(\ell)=G_{11}(-3) &\stackrel{!}{=} M\frac{\pi}{28} \int_{k\leq \Lambda} \frac{\mbox{d}^2k}{(2\pi)^2} \frac{\tanh(\frac{\pi k}{2}) }{k^2+1}.
\end{align}
As a function of the number $\ell$ of rings  we have:
\begin{center}
\begin{tabular}{|c|c|c|c|c|c|c|}
\hline $\ell$ & 1 & 2 & 3 & 4 & 5 & 6 \\ 
\hline $C=G_{11}(-3)$ & 0.448 & 0.649 & 0.707 & 0.728 & 0.736 & 0.739 \\ 
 \hline $\Lambda$ & 4.59 & 8.87 & 10.7 & 11.4 & 11.7 & 11.9 \\
\hline $\Lambda_1$ & 4.06 & 7.89 & 9.52 & 10.2 & 10.5 & 10.6\\
\hline 
\end{tabular} 
\end{center}
We evaluate the integral and find
\begin{align}
C &= \frac{M}{56} \int_0^\Lambda \mbox{d}k\ k\ \frac{\tanh(\frac{\pi k}{2}) }{k^2+1},
\end{align}
which determines $\Lambda$.
If we neglect the tanh term, then
\begin{align}
C \approx \frac{M}{56} \int_0^{\Lambda_1} \mbox{d}k\ k\ \frac{1}{k^2+1} = \frac{M}{112} \log(1+\Lambda_1^2),
\end{align}
which defines
\begin{align}
 \Lambda_1 = \sqrt{e^{112C/M}-1} \simeq e^{56C/M}.
\end{align}
Hence the generic value for $\Lambda\simeq \Lambda_1$ is 10, and we have $\sqrt{M}=4.2$, and so
\begin{align}
 \Lambda \gtrsim 2 \sqrt{M}.
\end{align}

\section{Curvature-limited correlations \label{sec:limited}}

In this section, we derive Eq.~(\ref{xiLabel}) for the curvature-limited correlations in the limit $L=1$. As shown in Ref. \cite{Boettcher2020}, the correlation function $G(z,z,'\omega)$ for $L=1$ only depends on the hyperbolic distance $d(z,z')$. It is given by
\begin{align}
    \nonumber G(z,z',\omega) =  \frac{1}{2\pi}\text{Re}\Bigl[{} %
    Q_\nu\Bigl(\cosh(d(z,z')/\kappa)\Bigr) %
    -\mathcal{C}(\omega)\ P_\nu\Bigl(\cosh(d(z,z')/\kappa)\Bigr)\Bigr],
\end{align}
where $P_\nu$ ($Q_\nu$) are Legendre functions of the first (second) kind, $\nu=\frac{1}{2}(-1+\rm{i}\sqrt{M(\omega+3-\frac{1}{M})})$, and $\mathcal{C}(\omega)$ is a constant. We use the fact that the lower band edge is located at $E_0=-3+\frac{1}{M}+\mathcal{O}(h^3)$ for $L=1$ and write $\nu=\frac{1}{2}(-1+\rm{i}\sqrt{M(\omega-E_0)})$. We neglect the $\mathcal{O}(h^3)$ corrections in the following. 

Let us first consider the special case $\omega=E_0$ and show that the correlation length remains finite. We have $\nu=-1/2$ and find that $\mathcal{C}(E_0)=-\rm{i}\pi/2$ is purely imaginary while $P_{-1/2}(y)$ is purely real. Hence the term $\mathcal{C} P_\nu$ does not contribute to the Green function for $\omega =E_0$. We have $\text {Re}\ Q_{-1/2}(x) \sim \pi/\sqrt{2x}$ for $x\to \infty$, or
\begin{align}
  \label{xi2}  \text{Re}\ Q_{-1/2}\Bigl(\cosh(d/\kappa)\Bigr) \sim \pi e^{-d/(2\kappa)}
\end{align}
as $d/\kappa\to \infty$. Consequently, for $\omega=E_0$, the correlation length satisfies $\xi(E_0)=\kappa$, which is finite. 

Now consider frequencies $\omega < E_0$ below the band edge. We observe that $\nu=-\frac{1}{2}(1+\sqrt{M(E_0-\omega)})$ is real and negative. The function $\text{Re}\ Q_\nu(x)$ scales like $x^\nu$ for $x\to \infty$. Hence
\begin{align}
 \label{xi3}   \text{Re}\ Q_{\nu}\Bigl(\cosh(d/\kappa)\Bigr) \approx %
 \cosh(d/\kappa)^\nu  \approx%
 e^{d \nu/\kappa},
\end{align}
or
\begin{align}
 \label{xi4}   \frac{1}{\xi(\omega)}  \approx %
 \frac{1}{\kappa}\Bigl( 1+\sqrt{M(E_0-\omega)}\Bigr).
\end{align}
We independently verified this scaling behavior numerically from fitting the long-distance decay of $G(r,0,\omega)$ for $\omega \lesssim E_0$.

\section{Bound states \label{sec:bs}}
In this section, we derive the bound-state equations for a single qubit and for two qubits---Eqs.~(6) and (12) in the main text.
\subsection{Bound state for a single qubit}
Working directly in the continuum and in Fourier space (valid for $L=1$), we can write down the single-excitation wavefunction
\begin{equation}
\label{eq:bound-state-wf-momentum}
\ket{\psi} =\qty[ \int_K u(K)a^\dagger(K)+c_{1}\sigma^+_1]\ket{\downarrow0},
\end{equation}
where $\int_K \equiv \int \frac{d^2k}{(2\pi)^2}\tanh(\frac{\pi k}{2})$.
Schroedinger's equation ($\hat{H}\ket{\psi_B}=E_B\ket{\psi_B}$) yields an equation for the bound-state energy: 
\begin{align}
E_B &= \Ezero + \Sigma(E_B),\\
\Sigma(E_B)&=\tilde{g}^2\int_{K}\frac{1}{E_B-\omega(k)}=-\frac{4\tilde{g}^2}{3th^2}G\qty(z,z;\lambda =\frac{4}{3th^2} (3t+E_B)),
\end{align}
and for the photonic component of the wavefunction 
\begin{equation}
u(K)=c_1\tilde{g}\frac{\psi_K^*(z_1)}{E_B-\omega(k)}.
\end{equation}
The remaining coefficient $c_1$ is determined by the normalization condition.

Using the Fourier transform relation in Eq.~\eqref{eq:Fourier-rels}, the bound-state wavefunction can also be written in real space and is given in Eq.~(6) of the main text.

\subsection{Two qubits}
For two qubits (at positions 1 and 2,  respectively), the single-excitation wavefunction can be written as 
\begin{equation}
	\ket{\psi} =\qty[ \int_K u(K)a^\dagger(K)+\sum_{i=1,2}c_{i}\sigma^+_i]\ket{\downarrow\downarrow0}.
\end{equation}
Schroedinger's equation can be written as 
\begin{equation}
	G^{-1}(E_B)\mqty(c_1\\c_2)=0,
\end{equation}
where the Green function is
\begin{equation}
\label{eq:2Spin-gf}
G(E_B) = \mqty(E_B-\omega_{q}-\Sigma_{11}(E_B)&-\Sigma_{12}(E_B)\\-\Sigma_{21}(E_B)&E_B-\omega_{q}-\Sigma_{22}(E_B))^{-1}
\end{equation}
and the self energies are
\begin{equation}
	\Sigma_{ij}(E_B) = \int_K\frac{\tilde{g}^2\psi_K(z_i)\psi_K^*(z_j)}{E_B-\omega(k)}=-\frac{4\tilde{g}^2}{3th^2}G\qty(z_i,z_j;\lambda =\frac{4}{3th^2} (3t+E_B)).
\end{equation}

Eq.~\eqref{eq:2Spin-gf} can be written as follows:
\begin{equation}
G(E_B) = \mqty(E_B-\omega_{q}-\Sigma_{11}(E_B)&-\Sigma_{12}(E_B)\\-\Sigma_{21}(E_B)&E_B-\omega_{q}-\Sigma_{22}(E_B))^{-1} = \qty[(E_B-\Ezero-\Sigma_{11}(E_B))\idop-\Sigma_{12}(E_B)\sigma_x]^{-1},
\end{equation}
where $\idop$ is the $2\times 2$ identity matrix and $\sigma_x$ is the Pauli x matrix.
The eigenstates are $\mqty(c_1\\c_2)=\mqty(1\\1)$ and $\mqty(c_1\\c_2)=\mqty(1\\-1)$. Hence, the bound-state energies $E_B^\pm$ of the symmetric and antisymmetric bound states are given by the poles of 
\begin{equation}
\frac{1}{E_B^\pm-\Ezero-\Sigma_{11}(E_B^\pm)\mp \Sigma_{12}(E_B^\pm)}.
\end{equation}

The wavefunctions are given by
\begin{equation}
\ket{\psi_{\pm}} = c_{\pm}\qty[ \tilde{g}^2\int_K \frac{\psi_K^*(z_1)\pm \psi_K^*(z_2)}{E_B^\pm-\omega(k)}a^\dagger(K)+(\sigma^+_1\pm\sigma^+_2)]\ket{\downarrow\downarrow0},
\end{equation}
where $c_\pm$ are normalization constants.
The real-space version is given in the main text, in Eq.~(12).

\section{Spontaneous emission and $j(\omega)$ \label{sec:emission}}
In this section, we derive the spontaneous emission rate of an excited qubit within the continuum approximation---Eq.~\eqref{eq:jL} from the main text.

We consider a single qubit and work in the continuum approximation in Fourier space (for $L=1$). For weak coupling $g$, the photons can be integrated out within a Born-Markov approximation, giving rise to a Lindblad master equation for the reduced density matrix of the qubit described by (neglecting the Lamb shift)
\begin{equation}
\partial_t \rho = -i\comm{\Ezero \ket{\up}\bra{\up}}{\rho}+\Gamma\sigma^-\rho\sigma^+-\frac{\Gamma}{2}\acomm{\sigma^+\sigma^-}{\rho}.
\end{equation}
The decay rate $\Gamma$ is given in terms of the spectral function as follows: \begin{equation}
\Gamma = \frac{j(\Delta)}{2},
\end{equation}
where the spectral function is defined as (this can also be obtained from Fermi's golden rule)
\begin{equation}
j(\omega)=\pi\sum_K\abs{g_K}^2\delta(\omega-\omega_k).
\end{equation}
Here, $\omega_k$ describes the dispersion of the photons, given in Eq.~\eqref{eq:wk-def}, whereas $g_K$ can be read off from Eq.~\eqref{eq:H-cont-K-space-final}, giving $g_K=g\sqrt{\frac{\pi}{28}}\psi_K(z)$.

Explicitly, we find
\begin{align}
j(\omega) &= \pi g^2\frac{\pi}{28}\int\frac{d^2k}{(2\pi)^2} \tanh(\frac{\pi k}{2})\abs{\psi_K}^2\delta(\omega-\omega_k)\\
&=
\pi g^2\frac{\pi}{28}\int_0^\infty\frac{dk}{2\pi}k \tanh(\frac{\pi k}{2})\delta(\omega-E_0+k^2/M)\\
&=
\pi g^2\frac{\pi}{28}\int_0^\infty\frac{dk}{2\pi} k\tanh(\frac{\pi k}{2})\frac{\delta(k-\sqrt{(\omega-E_0)M})}{2\sqrt{(\omega-E_0)M}/M}\\
&=
\frac{\pi}{28}g^2\tanh(\frac{\pi\sqrt{(\omega-E_0)M}}{2})\frac{M}{4}.
\end{align}

 \clearpage
\end{widetext}
\end{document}